# PLASMA PROCESSING OF SRF CAVITIES AT JEFFERSON LAB: EXPERIMENT RESULTS AND SIMULATION INSIGHT*

I. H. Senevirathne†, T. Powers, N. Raut
Thomas Jefferson National Accelerator Facility, Newport News, USA

## Abstract

Plasma processing of superconducting radio frequency (SRF) cavities has been an active research effort at Jefferson Lab (JLab) since 2019, aimed at enhancing cavity performance by removing hydrocarbon contaminants and reducing field emission. In this experiment, processing using argon-oxygen and helium-oxygen gas mixtures to find minimum ignition power at different cavity pressure was investigated. Ongoing simulations are contributing to a better understanding of the plasma surface interactions and the fundamental physics behind the process. These simulations, combined with experimental studies, guide the optimization of key parameters such as gas type, RF power, and pressure to ignite plasma using selected higher-order mode (HOM) frequencies. This paper presents experimental data from argon-oxygen and helium-oxygen gas mixture C75 and C100 cavity plasma ignition studies, as well as simulation results for the C100-type cavity based on the COMSOL model previously applied to the C75 cavity.

## INTRODUCTION

Plasma cleaning using noble gases like argon (Ar) or helium (He) in combination with oxygen ($O_2$) is a common surface treatment technique to remove organic contaminants [1]. When RF power is applied to an SRF cavity filled with the gas mixture, the oscillating electric field accelerates free electrons. These energetic electrons ionize the gas mixture, producing reactive species such as $Ar^+$ or $He^+$, along with highly reactive oxygen radicals (O) and ions ($O_2^+$, $O^+$ and $O^-$). Although Ar and He are inert gases and they do not directly participate in chemical reactions with the contaminants, they help sustain the plasma. Meanwhile, the reactive oxygen species chemically react with organic contaminants (hydrocarbons, for example), breaking C–H and C–C bonds to form volatile products like CO, $CO_2$, and $H_2O$. This phenomenon can be represented in a simple form: $C_xH_y + O^- \rightarrow CO_2 + H_2O$, where reactive oxygen atoms oxidize hydrocarbons. This chemical oxidation effectively cleans the surface at the molecular level.

At Jefferson lab, plasma processing efforts includes plasma treatment of multi-cell cavities, cavity pairs, and in-situ processing of cryomodules. A robust and repeatable plasma processing procedure has been developed for the C100 style cavities, resulting in significant energy gain, averaging around 2.7 MV/m per cavity [2]. Plasma processing has also been successfully optimized for the C75-type cavities, establishing reliable ignition.

* Work supported by the funding announcement DE-FOA-0003261
† iresha@jlab.org

## PLASMA IGNITION EXPERIMENT

We conducted a plasma ignition experiment using argon (Ar) helium (He) and oxygen ($O_2$) mixtures to determine the minimum RF power required to ignite a room temperature plasma in different cells of the C75 and C100 superconducting radio-frequency (SRF) cavities.

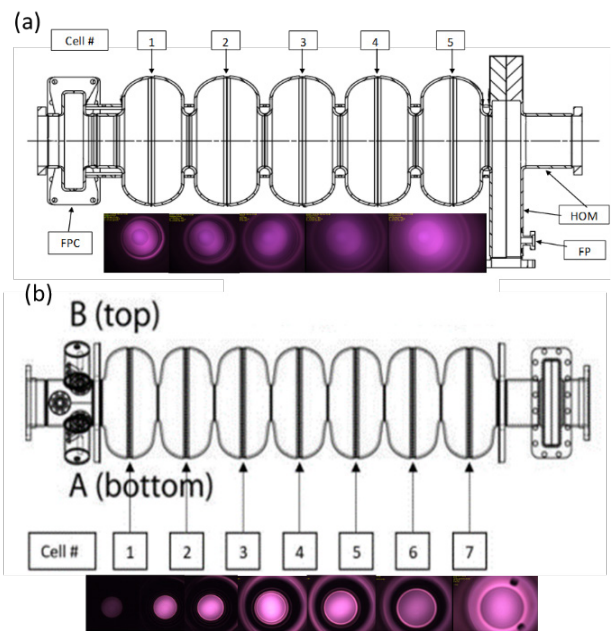

Figure 1: Cell numbering on (a) C75 cavity and (b) C100 cavity along with camera view of argon-oxygen plasma inside each cell.

S-parameters, particularly S11 and S21, are used to characterize the coupling and transmission behavior of electromagnetic modes in SRF cavities. S11 represents the reflection coefficient at the input port and provides insight into how efficiently a mode couples into the cavity, larger negative values indicate stronger coupling. S21, on the other hand, measures transmission through the cavity, helping identify whether specific modes, such as higher-order modes (HOMs), are propagating or suppressed. In plasma ignition experiments, HOMs, often dipole modes are commonly excited because they typically exhibit better coupling than the fundamental accelerating mode. Furthermore, the fundamental power coupler (FPC) setup for cryogenic operations of beam loaded cavity, therefore couples only minimal power into fundamental modes in room temperature. For plasma processing of C100-type cavities, TE111 HOMs in the 1850–2050 MHz range are used due to strong coupling via ILC-style HOM couplers, with 20 – 95% of applied power entering the cavity [3]. During plasma processing of the C100-type cavity, the input RF signal is fed through the HOM coupler, typically the bottom (A) one and exits via the FPC port. Cell counting begins from the cell nearest to the input (HOMs) as shown in Fig. 1 (b).

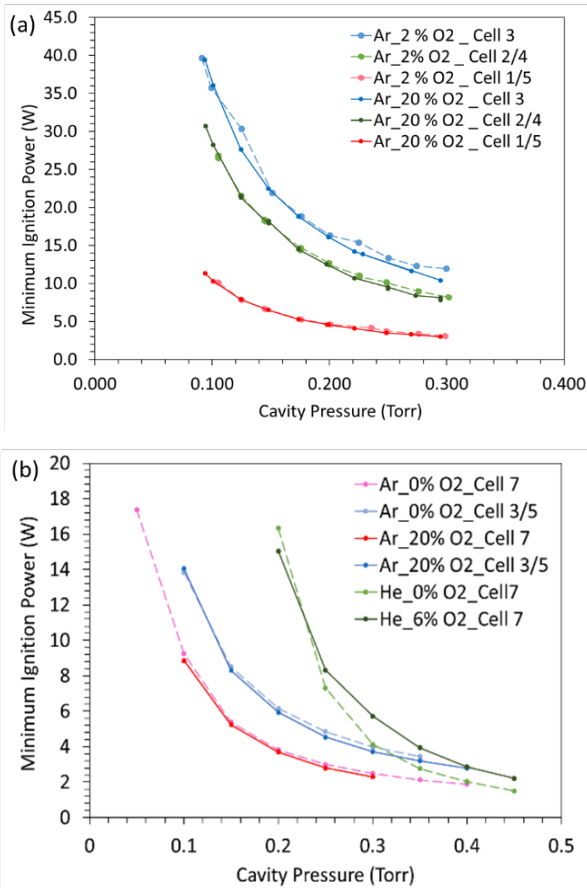

Figure 2: Minimum ignition power as a function of cavity pressure on (a) C75 (cell 3- 1713 MHz, cell 2/4- 1759 MHz and cell 1/5- 1827 MHz) and (b) C100 (cell7-1888 MHz and cell 3/5- 1940 MHz) cavity with different oxygen concentration.

For the C75 cavities, the FPC provides moderately effective coupling to the first three TE111 modes. Therefore, the input RF signal is fed through the FPC port during plasma processing. Cell counting begins from the cell nearest to the FPC, as shown in the Fig. 1 (a).

The room temperature measurements were performed to determine the minimum RF power required for plasma ignition on both C75 and C100 cavities. We applied RF energy at selected frequencies, chosen based on the electric field distribution, to ignite plasma in the desired cavity cell. The results show that higher frequencies tend to require lower ignition power as shown in Fig. 2. The minimum ignition power decreases gradually with increasing cavity pressure. Notably, the C75 cavity requires higher power compared to the C100 cavity. Experiments were performed at 2% and 20% $O_2$ concentrations for the C75 cavity, and at 0% and 20% $O_2$ for the C100 cavity. The data indicate that the oxygen concentration has no significant effect on the minimum power required to ignite the plasma when using Ar-$O_2$ mixture.

We also conducted similar experiments using helium-oxygen gas mixture using 0% and 6% $O_2$ concentration in C100 cavity. The results showed that the ignition power is slightly higher at elevated pressures when 6% oxygen is added, compared to pure helium. Currently, a helium gas mixture with 6% oxygen is used for plasma processing of CEBAF production cavities, as it offers ease of stabilization during the process. Our experiments demonstrate that helium is more effective than argon in achieving reliable plasma ignition and maintaining stable conditions for efficient cavity cleaning [4].

For CEBAF production cavity, plasma processing is performed using helium (94%)-oxygen (6%) mixture at 0.3 Torr and carefully increase the RF power to generate a dense plasma without causing coupler breakdown, thereby enhancing the cleaning efficiency.

## ONGOING SIMULATION

We have used COMSOL® Multiphysics simulation software [5] to model and analyse the plasma behaviour within our system. The software's Plasma Module enables the calculation of electron density within a plasma ignited in a favourable cell of an SRF cavity. By specifying operating parameters such as gas pressure and RF power, the model solves the coupled set of plasma physics and electromagnetic equations.

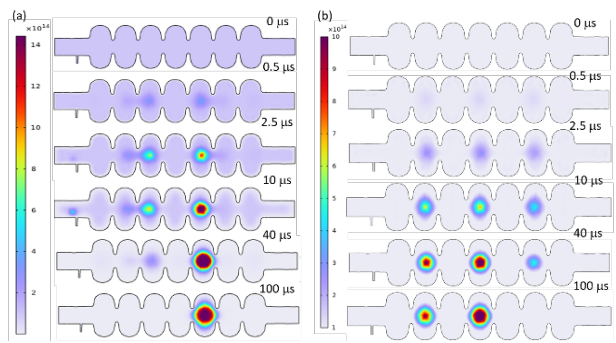

Figure 3: Ar-$O_2$ plasma ignition in C100 cavity with time on (a) cell 3/5, 1947 MHz and (b) cell 2/4/6, 1980 MHz in COMSOL simulation (ignition power 5 W, 6% $O_2$ and gas pressure 0.15 Torr).

In particular, our simulations applied to the C75 cavity showed that for an Ar-$O_2$ plasma, higher power (5-20 watts) was required to ignite the plasma for 20% oxygen, whereas lower power (2.5-10 watts) was required to ignite for 2% oxygen. For the 20% oxygen mixture, we observed growth in the electron number density (Ne) as a function of the pressure at 5 W and 10 W for all three modes. The Ne at powers 2.5 W and 5 W at 2% oxygen is higher than that at any power and pressure at 20% oxygen. Here, the lower power <7.5 W is good enough for plasma ignition. Hence, the 2% oxygen concentration can be nearly as effective as the 20% concentration at lower power levels [6]. It is also possible to study changes in the distribution of the number density of oxygen molecule such as O, $O^-$, $O^+$ and $O_2^+$ during plasma ignition within the cell, particularly along different spatial directions [7].

In this work, this model is applied to the C100-type cavity. In this case, we were able to perform 3D simulations under more realistic conditions. With the actual electromagnetic domain, the simulation time increased significantly, even when using a coarse mesh. Figure 3 illustrates the plasma ignition behavior at 1947 MHz and 1980 MHz in the C100 cavity over time, showing stable and localized ignition patterns. Figure 4 presents the corresponding electron density profiles at these frequencies along with ignition timing.

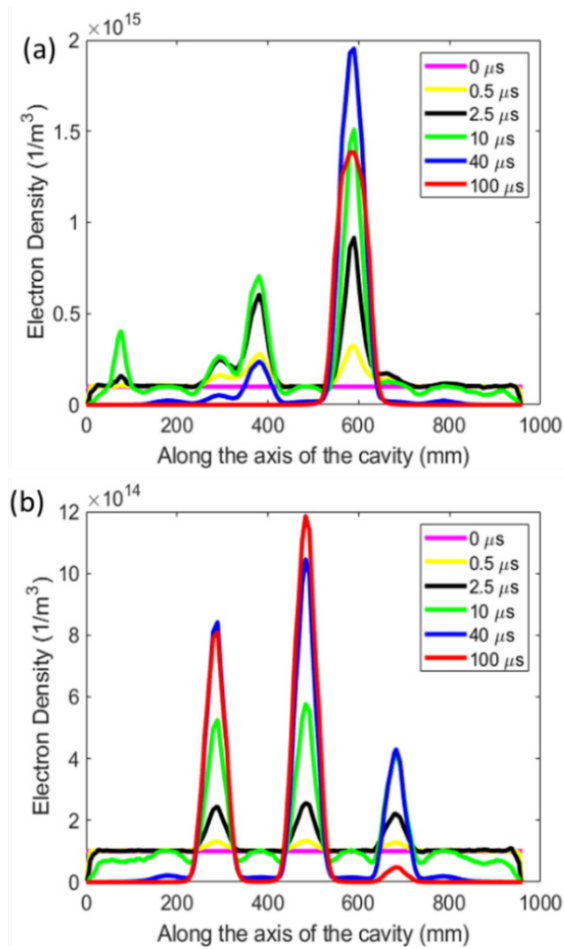

Figure 4: Electron density at plasma ignition with time (a) cell 3/5, 1947 MHz and (b) cell 2/4/6, 1980 MHz (ignition power 5 W, 6% O2 and gas pressure 0.15 Torr).

Plasma tends to ignite and persist in regions with the highest E-field strength, as these areas provide sufficient energy to ionize the gas and maintain the plasma state. At 1947 MHz, the electric field is more concentrated in cells 3 and 5, with a stronger preference toward cell 5, causing the plasma to eventually migrate entirely to cell 5. At 1980 MHz, the electric field is distributed across cells 2, 4, and 6, with plasma predominantly accumulating in cell 4, while a smaller portion remains in cell 2. In practice, we can use 1947 MHz to drive plasma into cells 3 and 5, and 1980 MHz to target cells 2, 4, and 6. Although initiating plasma in a specific cell can be challenging, once ignition occurs in cell 7, it becomes easier to move the plasma from one cell to another using appropriate frequencies.

Using Ar-$O_2$ based gas mixtures, neutral oxygen atoms (O) and oxygen ions ($O^+$, $O_2^+$, $O^-$) dominate the reactive species responsible for decomposing hydrocarbon residues on niobium surfaces. This simulation (see Fig. 5) indicates elevated atomic-O concentrations along both the cavity axis and wall where hydrocarbon removal is most efficient. Atomic oxygen, in particular, is identified as the key reactive species. It oxidizes hydrocarbon contaminants into volatile byproducts CO, $CO_2$, and $H_2O$ which are then removed through the flowing process gas. The effectiveness and selectivity of these removal reactions are critically dependent on electron temperature, plasma ignition power, cavity pressure, and the $O_2$ fraction in the gas mixture [8].

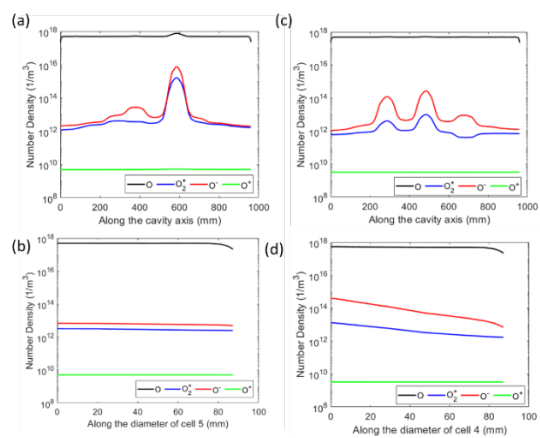

Figure 5: Different species of oxygen at ignition power 5 W, 6% O2 and gas pressure 0.15 Torr (a) on the axis of the cavity at 1947 MHz (b) along the diameter of cell 5 at 1947 MHz (c) on axis of the cavity at 1980 MHz and (d) along the diameter of cell 4 at 1980 MHz.

## CONCLUSION

We successfully extended our plasma ignition studies to both C75 and C100-type cavities. The results indicate that oxygen ($O_2$) concentration has minimal influence on the minimum RF power required for Ar-$O_2$ plasma ignition, whereas it plays a significant role in He-$O_2$ plasma. Building on our earlier success with C75 plasma simulations, we applied a 3D COMSOL model to the C100 cavity under more realistic conditions. This allowed us to map the electron density at the time of plasma ignition at 1947 MHz (cell 3/5) and 1980 MHz (cell 2/4/6). Furthermore, using Ar-$O_2$ gas mixture, we mapped the distribution of atomic oxygen (O) and oxygen ions ($O^+$, $O_2^+$, $O^-$) along the cavity axis and across the transverse axis of selected cells (cell 5 at 1947 MHz and cell 4 at 1980 MHz). These reactive species are primarily responsible for the removal of hydrocarbon residues from niobium surfaces, highlighting the effectiveness of the plasma cleaning process.

## ACKNOWLEDGEMENTS

The author would like to acknowledge the support and efforts of the technical staff from the SRF department for their assistance in assembling the setup.